\documentclass[12pt,a4paper]{article}
\pdfoutput=1
\usepackage[a4paper]{geometry}

\usepackage{amsmath}
\usepackage{amsthm}
\usepackage{amsfonts}
\usepackage{amssymb}
\usepackage[nosort]{cite}
\usepackage{hyperref}
\usepackage{multirow}
\usepackage{color}
\usepackage{xcolor}
\usepackage{mdframed}
\usepackage{tcolorbox}
\usepackage{framed}
\usepackage{tikz}
\usepackage{footmisc}

\newcommand{\op}{\hspace{1pt}}

\newcommand{\CC}{\mathbb{C}}
\newcommand{\RR}{\mathbb{R}}

\newtheorem*{nohslambdaa}{Hs[$\lambda$] is not compatible with $\mathfrak{v}\mathfrak{e}\mathfrak{c}\mathfrak{t}(S^2)$}

\newtheorem*{nolindef}{No linear deformations of $\mathfrak{v}\mathfrak{e}\mathfrak{c}\mathfrak{t}(S^2)$}

\newtheorem*{nochiraldef}{No linear deformations of chiral $A^{\pm}$ subalgebras}

\allowdisplaybreaks[2]
\numberwithin{equation}{section}

\begin{document}

\vspace*{-1.5cm}
\begin{flushright}
  {\small
  LMU-ASC 13/21
  }
\end{flushright}

\vspace{1.75cm}

\begin{center}
{\LARGE On deformations and extensions of 
 $\text{Diff}(S^2)$  \\ 
}
\end{center}

\vspace{0.4cm}

\begin{center}
  Mart\'in Enr\'iquez Rojo$^{1}$, Tom\'a\v{s} Proch\'azka$^{1,2}$, Ivo Sachs$^{1}$
\end{center}

\vspace{0.3cm}

\begin{center} 
\textit{$^{1}$\op Arnold Sommerfeld Center for Theoretical Physics\\[1pt]
Ludwig-Maximilians-Universit\"at \\[1pt]
Theresienstra\ss e 37 \\[1pt]
80333 M\"unchen, Germany}\\[1em]
\textit{$^{2}$\op Institute of Physics AS CR\\[1pt]
Na Slovance 2 \\[1pt]
Prague 8, Czech Republic}\\[1em]
\textit{Email:} \href{mailto:martin.enriquez@physik.lmu.de}{\texttt{martin.enriquez@physik.lmu.de}}, \\
\href{mailto:prochazkat@fzu.cz}{\texttt{prochazkat@fzu.cz}}, \\
\href{mailto:ivo.sachs@physik.lmu.de}{\texttt{ivo.sachs@physik.lmu.de}}
\end{center}

\vspace{1.8cm}


\begin{abstract}
\noindent
We investigate the algebra of vector fields on the sphere. First, we find that linear deformations of this algebra are obstructed under reasonable conditions. In particular, we show that $hs[\lambda]$, the one-parameter deformation of the algebra of area-preserving vector fields, does not extend to the entire algebra. Next, we study some non-central extensions through the embedding of $\mathfrak{v}\mathfrak{e}\mathfrak{c}\mathfrak{t}(S^2)$ into $\mathfrak{v}\mathfrak{e}\mathfrak{c}\mathfrak{t}(\CC^*)$. For the latter, we discuss a three parameter family of non-central extensions which contains the symmetry algebra of asymptotically flat and asymptotically Friedmann  spacetimes at future null infinity, admitting a simple free field realization. 

\end{abstract}


\clearpage

\tableofcontents


\section{Introduction}

The algebra of vector fields on the circle $\mathfrak{v}\mathfrak{e}\mathfrak{c}\mathfrak{t}(S^1)$ and its Virasoro central extension have played a major role  in quantum gravity and theoretical high energy physics for the last half century. It is the symmetry algebra of two dimensional conformal field theories (CFT) \cite{Belavin:1984vu}, plays a fundamental role in string theory and appears constantly in a wide range of applications, among which we would like to highlight black hole microstate counting \cite{Strominger:1997eq,Birmingham:2001dt}, fluid-gravity duality \cite{Penna:2017vms,Carrillo-Gonzalez:2018wrh}, and asymptotic symmetries in three dimensions \cite{Brown:1986nw,Seraj:2016cym,Oblak:2016eij}.

Nonetheless, it is the algebra of vector fields on the sphere that naturally arises when trying to describe two dimensional membranes, instead of strings, and to approach black hole microstate counting, fluid-gravity duality and asymptotic symmetries in four dimensions. 

In fact, it was shown that the spherical two dimensional membrane in light-cone gauge is invariant under area preserving diffeomorphisms on the sphere $\text{SDiff}(S^2)$, whose algebra of smooth vector fields is denoted by $\mathfrak{s}\mathfrak{v}\mathfrak{e}\mathfrak{c}\mathfrak{t}(S^2)$, and that their large $N$ $SU(N)$ discretization is presently the most viable path to membrane quantization \cite{1982PhDT........32H,deWit:1988wri,Nicolai:1998ic}. More precisely, a one parameter family of algebras, known as $hs[\lambda]$, reduces to $SU(N)$ for integer $\lambda$ and becomes $\mathfrak{s}\mathfrak{v}\mathfrak{e}\mathfrak{c}\mathfrak{t}(S^2)$ in the limit $\lambda\to\infty$ \cite{Pope:1990kc} \footnote{It is worth noting that $\text{SDiff}(S^2)$ and its deformation $hs[\lambda]$ also play an important role in
the context of higher spin $\text{AdS}_3/\text{CFT}_2$ correspondence \cite{Gaberdiel:2012uj,Perlmutter:2012ds} and, particularly, in the structure and properties of $W_\infty$-algebras \cite{Prochazka:2014gqa}.}.

Furthermore, non-central extensions of the algebra of vector fields on the sphere have been recently considered to correspond to asymptotic symmetry algebras of asymptotically flat, $\mathfrak{g}\mathfrak{b}\mathfrak{m}\mathfrak{s}$ \cite{Campiglia:2014yka,Campiglia:2015yka,Compere:2018ylh,Donnelly:2020xgu,Freidel:2021yqe,Freidel:2021cbc}, and asymptotically decelerating spatially flat Friedmann-Lema\^itre-Robertson-Walker (FLRW), $\mathfrak{g}\mathfrak{b}\mathfrak{m}\mathfrak{s}_s$ \cite{Enriquez-Rojo:2021blc}, spacetimes at future null infinity, and asymptotically (anti) de-Sitter \cite{Compere:2019bua} in four spacetime dimensions. Non-central extensions of this algebra have also been discussed in the context of asymptotic symmetries in null hypersurfaces (including event horizons) \cite{Chandrasekaran:2018aop,Flanagan:2019vbl}.  Moreover, by means of the membrane paradigm \cite{Thorne:1986iy,Parikh:1997ma}, a connection between these asymptotic symmetries at null hypersurfaces and fluids on the sphere has been elucidated in \cite{Penna:2015gza,Penna:2017bdn}.

Thus, the study of the algebra of vector fields on the sphere is in order to deepen into the aforementioned physical research fields. Rather surprisingly, few studies have been performed trying to investigate the structure and properties of this algebra, as far as we are aware. In \cite{Bars:1988uj}, it has been shown that $\mathfrak{s}\mathfrak{v}\mathfrak{e}\mathfrak{c}\mathfrak{t}(S^2)$ does not admit central extensions, while \cite{Frappat:1989gn} studied generalized Kac-Moody algebras as loop algebras for $\mathfrak{v}\mathfrak{e}\mathfrak{c}\mathfrak{t}(S^2)$, and \cite{Saidi:1990px} investigated harmonic distributions on the sphere and related them to $\text{Diff}(S^2)$. More recently, the representation theory of $\text{SDiff}(S^2)$ has been explored using the method of coadjoint orbits in \cite{Penna:2018bzj}.

In this paper, we further investigate this algebra following two main paths. First, we analyze the structure and deformations of the algebra of globally defined vector fields on the sphere, $\mathfrak{v}\mathfrak{e}\mathfrak{c}\mathfrak{t}(S^2)$ as well as its ``chiral" subalgebra generated by holomorphic and anti-holomorphic vector fields respectively. Next, we embed $\mathfrak{v}\mathfrak{e}\mathfrak{c}\mathfrak{t}(S^2)$ in the algebra of vector fields on the two-punctured sphere, or punctured complex plane $\mathfrak{v}\mathfrak{e}\mathfrak{c}\mathfrak{t}(\CC^*)$, in order to investigate some of its physically relevant non-central extensions and devising simple free field realizations for them. Our main findings are:

\begin{enumerate}
    \item The chiral subalgebras contain half-Witt subalgebras generated by smooth vector fields on $S^2$. Furthermore, these chiral subalgebras can be reconstructed from a half-Witt subalgebra and horizontal operators as described in figure \ref{Agen}. 
    \item We present an argument for the absence of linear deformations of the algebra of smooth diffeomorphisms on the sphere, $\mathfrak{v}\mathfrak{e}\mathfrak{c}\mathfrak{t}(S^2)$. In particular, we show that the deformation family $hs[\lambda]$ \cite{Pope:1990kc} does not extend from $\mathfrak{s}\mathfrak{v}\mathfrak{e}\mathfrak{c}\mathfrak{t}(S^2)$ to $\mathfrak{v}\mathfrak{e}\mathfrak{c}\mathfrak{t}(S^2)$.
   
    \item In terms of the locally defined vector fields on the two-punctured sphere, we describe a three parameter family of non-central extensions $gW(a,b,\bar{a})$ which contains $\mathfrak{g}\mathfrak{b}\mathfrak{m}\mathfrak{s}$ \cite{Campiglia:2014yka,Campiglia:2015yka} and $\mathfrak{g}\mathfrak{b}\mathfrak{m}\mathfrak{s}_s$ \cite{Enriquez-Rojo:2021blc}. It generalizes and includes the $W(a,b;\bar{a},\bar{b})$ family of deformations for $\mathfrak{b}\mathfrak{m}\mathfrak{s}$ \cite{Safari:2019zmc,Safari:2020pje}, and can be realized by a simple free field realization. In addition, following the fact that $W(a,b;\bar{a},\bar{b})$ admits a central extension, $\hat{W}(a,b;\bar{a},\bar{b})$, we obtain an equivalent extension for $gW(a,b,\bar{a})$, which we name $\hat{gW}(a,b,\bar{a})$.
\end{enumerate}

This paper is organized as follows: in section \ref{vectS2}, we describe $\mathfrak{v}\mathfrak{e}\mathfrak{c}\mathfrak{t}(S^2)$ in two different bases that are adapted to the deformation problem of $\mathfrak{v}\mathfrak{e}\mathfrak{c}\mathfrak{t}(S^2)$ and its chiral subalgebras. We then investigate its linear deformations with the help of an attached Mathematica file \textsc{vectS2deformations.nb} \cite{Mathematica}. In section \ref{lvectS2}, we discuss the structure of $\mathfrak{v}\mathfrak{e}\mathfrak{c}\mathfrak{t}(\CC^*)$ and study some of its non-central extensions present in the context of asymptotic symmetries. We conclude with a summary of results and future endeavours in section \ref{conclusion}.

\section{Vector fields on $S^2$} 
\label{vectS2}

We begin by reviewing some basic properties of the algebra of smooth vector fields on the sphere $\mathfrak{v}\mathfrak{e}\mathfrak{c}\mathfrak{t}(S^2)$. In the following subsection, we then address the problem of linear deformations of this algebra.

\subsection{Description of the classical algebra}

The generators of $\mathfrak{v}\mathfrak{e}\mathfrak{c}\mathfrak{t}(S^2)$ fall  naturally into two classes \footnote{We use the conventions for the spherical harmonics $Y^\ell_m$ in Mathematica \cite{Mathematica} up to a global prefactor in order to have standard normalization for the $so(1,3)$ subalgebra.} 
\begin{itemize}
    \item Area-preserving
    \begin{eqnarray}
T^\ell_m&=&i\sqrt{\frac{4\pi}{3}}\epsilon^{ab}(\partial_b Y^\ell_m)\partial_a=\frac{i}{\text{sin}\theta}\sqrt{\frac{4\pi}{3}}\left((\partial_{\varphi}Y^\ell_m)\partial_{\theta}-(\partial_{\theta}Y^\ell_m)\partial_{\varphi}\right)
\end{eqnarray}
    \item Non area-preserving
    \begin{eqnarray}
S^\ell_m&=&i\sqrt{\frac{4\pi}{3}}g^{ab}(\partial_aY^\ell_m)\partial_b=i\sqrt{\frac{4\pi}{3}}\left((\partial_{\theta}Y^\ell_m)\partial_{\theta}+\frac{1}{\text{sin}^2\theta}(\partial_{\varphi}Y^\ell_m)\partial_{\varphi}\right)  
\end{eqnarray}
\end{itemize}
where $l>0$ and $-l\leq m \leq l$ denote the orbital and magnetic quantum numbers respectively, while  $\epsilon^{\theta\varphi}=\frac{1}{\sin(\theta)}$ and $g^{ab}$ are the inverse volume form and metric of the round sphere respectively. Let us now summarize some features of this algebra that will be useful in the sequel:

\begin{enumerate}
    \item The area preserving vector fields $T^l_m$ form a closed subalgebra called $\mathfrak{s}\mathfrak{v}\mathfrak{e}\mathfrak{c}\mathfrak{t}(S^2)$ and  \{$T^l_0$\} form an abelian closed subalgebra of the latter \footnote{There are infinitely many subalgebras of this form corresponding to different choices of $z$ axis.}. On the other hand, the non-area preserving vector fields $S^l_m$ do not close on themselves.
    \item The generators with $l=1$ form a subalgebra. In particular, 

\begin{equation}
    L_3:=-T^1_0 \ , \ \ L_{1}:=\frac{1}{\sqrt{2}}(T^{1}_{1}-T^{1}_{-1}) \ , \ \ L_{2}:=\frac{1}{\sqrt{2}i}(T^{1}_{1}+T^{1}_{-1}) \ ,
\end{equation}
with 
\begin{equation}
    [L_i,L_j]=i\epsilon_{ijk}L_k \ \ i,j=1,2,3 \ .
    \label{eq:1vect}
\end{equation}
generate the $so(3)$ subalgebra of rotations. Together with  
\begin{equation}
    S_3:=-S^1_0 \ , \ \ S_{1}:=\frac{1}{\sqrt{2}}(S^{1}_{1}-S^{1}_{-1}) \ , \ \ S_{2}:=\frac{1}{\sqrt{2}i}(S^{1}_{1}+S^{1}_{-1}) 
\end{equation}
and the commutation relations 
\begin{align}
    [L_i,S_j]&=[S_i,L_j]=i\epsilon_{ijk}S_k \\
        \label{eq:2vect}
    [S_i,S_j]&=-i\epsilon_{ijk}L_k \nonumber
\end{align}
they generate the subalgebra $so(1,3)$ of conformal diffeomorphisms on the sphere. 

\item The Lie algebra isomorphism $so(1,3)\simeq sl(2,\RR)\oplus sl(2,\RR)$ is made manifest by the complex linear combinations, $A^{\pm}_i:=\frac{1}{2}(L_i\pm iS_i)$, with
\begin{eqnarray}
\left[A^+_i,A^+_j\right]=i\epsilon_{ijk}A^+_k  \ , \ \ \ \
\left[A^-_i,A^-_j\right]=i\epsilon_{ijk}A^-_k  \ , \ \ \ \ 
\left[A^+_i,A^-_j\right]=0 \ .
\label{eq:splitting}
\end{eqnarray}

\item The generators with $\ell>1$ transform as vectors under the $so(3)$ subalgebra of rotations, that is ($B^{l}_m\in\{T^l_m,S^l_m\}$ ) 
\begin{align}
    [T^1_0,B^l_m]=-mB^{l}_m \nonumber\\
    [T^1_1,B^l_m]=\frac{\sqrt{(l+m+1)(l-m)}}{\sqrt{2}}B^{l}_{m+1} \\
    [T^1_{-1},B^l_m]=-\frac{\sqrt{(l-m+1)(l+m)}}{\sqrt{2}}B^{l}_{m-1} \ .\nonumber
\end{align}
In addition, they transform in a representation of $so(1,3)$ but, since the latter is infinite dimensional, this will not be of use here.
\item They have a definite transformation under parity,  $\mathcal{P}:\;\theta\to\pi-\theta$, $\phi\to\pi+\phi$, with $\mathcal{P}(T^l_m)=(-1)^{l+1}$, $\mathcal{P}(S^l_m)=(-1)^{l}$. The commutation relations are compatible with parity. 
\end{enumerate}

\subsubsection{Chiral subalgebras}
The decomposition of the algebra as in (\ref{eq:splitting}) does not generalise to $\ell>1$. However, there are subalgebras $A^\pm$ for $\ell>1$. This is more easily seen in  stereographic coordinates 
\begin{equation}
    z=e^{i\varphi}\text{cot}(\theta/2) \ , \ \ \ \bar{z}=z^* \ .
\end{equation}
In these coordinates we have 
\begin{equation}
    T^l_m=\sqrt{\frac{4\pi}{3}}[\frac{(1+z\bar{z})^2}{2}[(\partial_z Y^l_m)\partial_{\bar{z}}-(\partial_{\bar{z}} Y^l_m)\partial_{z}] \ ,
\end{equation}
\begin{equation}
    S^l_m=i\sqrt{\frac{4\pi}{3}}\frac{(1+z\bar{z})^2}{2}[(\partial_z Y^l_m)\partial_{\bar{z}}+(\partial_{\bar{z}} Y^l_m)\partial_{z}] \ ,
\end{equation}
which makes the decomposition of  $\mathfrak{v}\mathfrak{e}\mathfrak{c}\mathfrak{t}(S^2)$ into  holomorphic- and anti holomorphic vector fields manifest, that is 
\begin{align}
    (A^l_m)^+=-\sqrt{\frac{4\pi}{3}}\frac{(1+z\bar{z})^2}{2}(\partial_{\bar{z}} Y^l_m)\partial_{z} \ , \ \ (A^l_m)^-=\sqrt{\frac{4\pi}{3}}\frac{(1+z\bar{z})^2}{2}(\partial_z Y^l_m)\partial_{\bar{z}} \ . \label{eq:agen}
\end{align}
This reveals further subalgebras. Here we list some of their features:

\begin{enumerate}

  \item The $l=1$ subalgebra (\ref{eq:splitting}) is  recovered with 
\begin{eqnarray}
(A^{1}_{0})^+=-z\partial_z \ , \ \ \ \ 
(A^{1}_{1})^+=-\frac{1}{\sqrt{2}}z^2\partial_z \ , \ \ \ \ 
(A^{1}_{-1})^+=-\frac{1}{\sqrt{2}}\partial_z \ , \nonumber\\
(A^{1}_{0})^-=\bar{z}\partial_{\bar{z}} \ , \ \ \ \ 
(A^{1}_{1})^-=-\frac{1}{\sqrt{2}}\partial_{\bar{z}} \ , \ \ \ \ 
(A^{1}_{-1})^-=-\frac{1}{\sqrt{2}}\bar{z}^2\partial_{\bar{z}} \ .
\end{eqnarray}
Furthermore, $(A^l_m)^{\pm}$ transform as vectors under the $so(3)$ subalgebra of rotations
    \begin{align}
    [T^1_0,(A^l_m)^{\pm}]=-m(A^l_m)^{\pm} \nonumber \\
    [T^1_1,(A^l_m)^{\pm}]=\frac{\sqrt{(l+m+1)(l-m)}}{\sqrt{2}}(A^l_{m+1})^{\pm} \\
    [T^1_{-1},(A^l_m)^{\pm}]=-\frac{\sqrt{(l-m+1)(l+m)}}{\sqrt{2}}(A^l_{m-1})^{\pm} \ . \nonumber
    \end{align}

    \item The chiral algebras $A^{\pm}=\{(A^l_m)^{\pm}\}$ form subalgebras, mapped to each other by parity and which do not commute, $[(A^l_m)^{+},(A^{l'}_{m'})^{-}]\neq0$ for $l,l'>1$. They can be extended further by $A^{\pm}\oplus (A^1_m)^{\mp}$ to maximal subalgebras\footnote{Orthogonality is with respect to the canonical inner product on $S^2$.}. In addition to $\{(A^1_m)^{\mp}\}$, both of the latter admit $\{A^l_0\}$ as non-abelian subalgebras. Further  subalgebras are generated by $\{(A^{\ell}_{\pm\ell})^+\}$,  $\{(A^{\ell}_{\pm\ell})^-\}$, as well as  $\{(A^l_{l})^+\}\cup \{(A^{l'}_{l'})^-\}$, $\{(A^l_{-l})^+\}\cup \{(A^{l'}_{-l'})^-\}$ and $\{(A^l_{0})^+\}\cup \{(A^{l'}_{0})^-\}$.  
    
     The subalgebras generated by $\{(A^{\ell}_{\pm\ell})^+\}$ (and similarly by $\{(A^{\ell}_{\pm\ell})^-\}$) are isomorphic to the subalgebra of the Witt algebra, $[L_n,L_m]=(m-n)L_{m+n}$, generated by $L_n$ with $n>0$, usually called half-Witt algebra. This can be seen by a change of normalization, for instance 
  {\scriptsize  
\begin{align}
    L_1=(A^1_1)^+ \ , \ 
    L_2=\frac{1}{2 \sqrt{10}}(A^2_2)^+ \ , \ 
L_3=\frac{1}{2 \sqrt{210}}(A^3_3)^+ \ , \ 
L_4=\frac{1}{8 \sqrt{210}}(A^4_4)^+ \ , \ 
L_5=\frac{1}{20 \sqrt{462}}(A^5_5)^+ \ . \nonumber
\end{align} }These half-Witt subalgebras miss a lowering operator and, contrary to the usual two dimensional conformal field theory on the sphere, the corresponding vector fields are regular everywhere.\footnote{Singular vector fields arise e.g. on the celestial sphere as in e.g. \cite{Ball:2019atb,Donnay:2020guq}.} 
    
     \item More generally, the generators of the chiral subalgebra $A^+$ (and similarly $A^-$) can be constructed from a single generator $(A^2_{-2})^+$ with a raising operator, $A^1_{-1}$ and a horizontal operator, $T^1_{\pm1}$, as described in fig. \ref{Agen}. 
   
    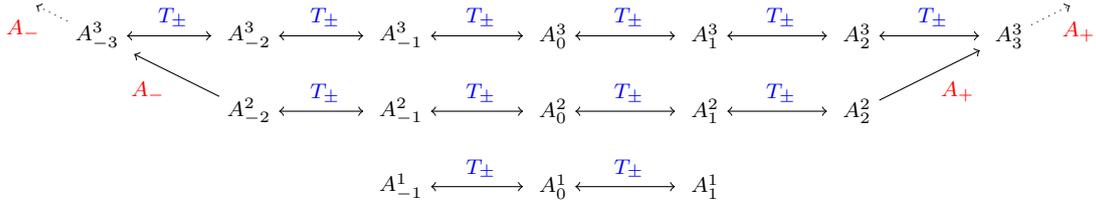
\begin{figure}[h!]
    \centering
{\scriptsize 
\begin{tikzpicture}
\node (1) at (5, -1) {$A^1_{-1}$};
\node (2) at (7, -1) {$A^1_{0}$};
\node (3) at (9, -1) {$A^1_{1}$};
\node (a1) at (3, 0) {$A^2_{-2}$};
\node (a2) at (5, 0) {$A^2_{-1}$};
\node (a3) at (7, 0) {$A^2_{0}$};
\node (a4) at (9, 0) {$A^2_{1}$};
\node (a5) at (11, 0) {$A^2_{2}$};
\node (b) at (1, 1) {$A^3_{-3}$};
\node (b2) at (3, 1) {$A^3_{-2}$};
\node (b3) at (5, 1) {$A^3_{-1}$};
\node (b4) at (7, 1) {$A^3_{0}$};
\node (b5) at (9, 1) {$A^3_{1}$};
\node (b6) at (11, 1) {$A^3_{2}$};
\node (b7) at (13, 1) {$A^3_{3}$};
\node (c) at (0, 1.5) {};
\node (c9) at (14, 1.5) {};
\draw[->, shorten > = 3pt] (a1) -- (b) node[midway, below left] { $\textcolor{red}{A_{-}}$};
\draw[dotted,->, shorten > = 3pt] (b) -- (c) node[midway, below left] { $\textcolor{red}{A_{-}}$};
\draw[->, shorten > = 3pt] (a5) -- (b7) node[midway, below right] { $\textcolor{red}{A_{+}}$};
\draw[dotted,->, shorten > = 3pt] (b7) -- (c9) node[midway, below right] { $\textcolor{red}{A_{+}}$};
\draw[<->, shorten > = 3pt] (a1) -- (a2) node[midway, above] { $\textcolor{blue}{T_{\pm}}$};
\draw[<->, shorten > = 3pt] (a2) -- (a3) node[midway, above] { $\textcolor{blue}{T_{\pm}}$};
\draw[<->, shorten > = 3pt] (a3) -- (a4) node[midway, above] { $\textcolor{blue}{T_{\pm}}$};
\draw[<->, shorten > = 3pt] (a4) -- (a5) node[midway, above] { $\textcolor{blue}{T_{\pm}}$};
\draw[<->, shorten > = 3pt] (b) -- (b2) node[midway, above] { $\textcolor{blue}{T_{\pm}}$};
\draw[<->, shorten > = 3pt] (b2) -- (b3) node[midway, above] { $\textcolor{blue}{T_{\pm}}$};
\draw[<->, shorten > = 3pt] (b3) -- (b4) node[midway, above] { $\textcolor{blue}{T_{\pm}}$};
\draw[<->, shorten > = 3pt] (b4) -- (b5) node[midway, above] { $\textcolor{blue}{T_{\pm}}$};
\draw[<->, shorten > = 3pt] (b5) -- (b6) node[midway, above] { $\textcolor{blue}{T_{\pm}}$};
\draw[<->, shorten > = 3pt] (b6) -- (b7) node[midway, above] { $\textcolor{blue}{T_{\pm}}$};
\draw[<->, shorten > = 3pt] (1) -- (2) node[midway, above] { $\textcolor{blue}{T_{\pm}}$};
\draw[<->, shorten > = 3pt] (2) -- (3) node[midway, above] { $\textcolor{blue}{T_{\pm}}$};
\end{tikzpicture}}
    \caption{Generation of chiral generators starting from $A^2_{\mp 2}$ and then first acting repeatedly with $A_{\mp}=A^1_{\mp1}$ to obtain $A^\ell_{\mp\ell}$ and then acting with $T_{\pm}=T^1_{\pm 1}$ to obtain $A^\ell_m$. }
    \label{Agen}
\end{figure}
\end{enumerate}
    Schematically we can understand the algebra as $\mathfrak{v}\mathfrak{e}\mathfrak{c}\mathfrak{t}(S^2)\simeq (so(3)\hookrightarrow hW) \cup (\overline{so(3)\hookrightarrow hW})$, where $so(3)\hookrightarrow hW$ denotes the action of $so(3)$ on half-Witt and the bar on top signals parity conjugation. This structure resembles and contains $so(1,3)\simeq sl(2,\mathbb{R}) \oplus \overline{sl(2,\mathbb{R})}$. 
\subsection{Linear deformations of $\mathfrak{v}\mathfrak{e}\mathfrak{c}\mathfrak{t}(S^2)$}

In this section, we investigate the linear deformations of $\mathfrak{v}\mathfrak{e}\mathfrak{c}\mathfrak{t}(S^2)$ and its chiral subalgebras $A^{\pm}$. By linear we mean that the commutation relations do not generate  higher powers of the generators. This problem is hard to tackle analytically because of the complicated form of the structure constants. Here, we reformulate the problem in a way that can be analyzed level by level with the help of Mathematica. 

To study deformations, we first have to specify the conditions that we impose on any consistent deformation. Concretely we impose:

\begin{enumerate}
    \item The Jacobi identities have to be satisfied.
    \item The generators ($T^j_m$, $S^j_m$) have to transform as spherical tensors under $T^1_m$ (i.e. the isometry group of $S^2$ is not deformed).
    \item The possible deformations have to include the classical algebra $\mathfrak{v}\mathfrak{e}\mathfrak{c}\mathfrak{t}(S^2)$ as a limit in the deformation parameter.
     \item The generators are required to have a definite transformation under parity. 
\end{enumerate}

The general ansatz for the commutators, imposing covariance under the rotation group (condition 2) reads as follows:
\begin{align}
[T^{j_1}_{m_1}, T^{j_{2}}_{m_2}]=&\sum^{j_{1}+j_{2}}_{j=|j_1-j_2|}[A(j_1,j_2,j)T^{j}_{m_1+m_2}+B(j_1,j_2,j)S^{j}_{m_1+m_2}]\times C^{j_1j_2j}_{m_1m_2m} \label{eq:ttcom} \\
[T^{j_1}_{m_1}, S^{j_{2}}_{m_2}]=&\sum^{j_{1}+j_{2}}_{j=|j_1-j_2|}[C(j_1,j_2,j)T^{j}_{m_1+m_2}+D(j_1,j_2,j)S^{j}_{m_1+m_2}]\times C^{j_1j_2j}_{m_1m_2m} \label{eq:tscom} \\
[S^{j_1}_{m_1}, S^{j_{2}}_{m_2}]=&\sum^{j_{1}+j_{2}}_{j=|j_1-j_2|}[E(j_1,j_2,j)T^{j}_{m_1+m_2}+F(j_1,j_2,j)S^{j}_{m_1+m_2}]\times C^{j_1j_2j}_{m_1m_2m} \label{eq:sscom} 
\end{align}
with the $m$-dependence of the commutators completely fixed by the $so(3)$ subalgebra (second condition) together with the Wigner-Eckart theorem such that, using the conventions of \cite{Pope:1990kc}:
{\footnotesize
$$C^{j_1j_2j}_{m_1m_2m}=\sum^{j_1+j_2-j}_{m=0} \binom{j_1+j_2-j}{m}[j_1-m_1]_m[j_1+m_1]_{j_1+j_2-j-m}[j_2-m_2]_{j_1+j_2-j-m}[j_2+m_2]_{m},$$}where the the combinatorial factor $[a]_n=a(a-1)...(a-n+1)$ is a Pochhammer symbol. 

There are infinitely many free coefficients in the above commutators. These will be reduced by imposing parity invariance and the Jacobi identity, while taking care at the same time that the solutions remain in the classical branch. Furthermore, for each $j$ there is a rescaling freedom of the generators ($T^j_m$, $S^j_m$) which we fix by choosing coefficients that do not vanish in $\mathfrak{v}\mathfrak{e}\mathfrak{c}\mathfrak{t}(S^2)$ and assign them a value. The practical way to perform this analysis at the computational level is to solve the Jacobi identities order by order in $j_{\text{max}}$ \footnote{For given value of $j_{\text{max}}$, we solve the Jacobi identities $[B^{j_1}_{m_1},[B^{j_2}_{m_2},B^{j_3}_{m_3}]]+\text{cyclic permutations}=0$ for $0\leq j_1\leq j_{\text{max}}$, $0\leq j_2\leq j_{\text{max}}-j_1$ and $0\leq j_3\leq j_{\text{max}}-j_1-j_2$, where $B^j_m$ represents both $T^j_m$ and $S^j_m$.}, replacing the coefficients of lower $j$ in terms of the ones with higher $j$ and analyze the resultant algebra at every level to observe if there are free parameters left. 

In \textsc{vectS2deformations.nb}, attached to this note, we carry out this analysis up to $j_{\text{max}}=7$ and we observe that the algebra of commutators with $j\leq 3$ is completely determined. In fact, it is clear from the computations that the number of independent equations grows faster than the number of free coefficients. It is not possible for us to continue this analysis to large $j_{\text{max}}$ due to lack of computational power. Nevertheless, from the computational perspective it is rather evident that the algebra will not admit deformations at higher $j$. On this ground, we arrive at the following claim:

\begin{nolindef} 

\begin{itemize}

    \item[] ~\\ The algebra of smooth diffeomorphisms on the sphere, $\mathfrak{v}\mathfrak{e}\mathfrak{c}\mathfrak{t}(S^2)$, does not admit linear deformations satisfying the Jacobi identities, parity and vector representation of the generators under rotations \footnote{Note that parity seems to follow naturally from the Jacobi identities, so it might actually not be a necessary requirement.\label{reqs}}.

\end{itemize}

\label{theorem:nodef}
\end{nolindef}

Let us contrast this result to the well known one parameter linear deformation of $\mathfrak{s}\mathfrak{v}\mathfrak{e}\mathfrak{c}\mathfrak{t}(S^2)$, which is also known as higher spin algebra or $hs[\lambda]$ \cite{Pope:1990kc}. The latter  is obtained from the so-called lone-star product of area-preserving generators \cite{Pope:1990kc}: 
{\footnotesize
\begin{eqnarray}
T^{j_1}_{m_1}\star T^{j_{2}}_{m_2}=\sum^{j_{1}+j_{2}}_{j=|j_1-j_2|}T^{j}_{m_1+m_2}\frac{1}{4^{j_1+j_2-j}(j_1+j_2-j)!}{}_{4}F_{3}\binom{\frac{1}{2}+\lambda, \frac{1}{2}-\lambda, \frac{1+j-j_1-j_2}{2}, \frac{j-j_1-j_2}{2}}{\frac{1}{2}-j_1, \frac{1}{2}-j_2,\frac{3}{2}+j} \nonumber \\
\times \sum^{j_1+j_2-j}_{m=0} \binom{j_1+j_2-j}{m}[j_1-m_1]_m[j_1+m_1]_{j_1+j_2-j-m}[j_2-m_2]_{j_1+j_2-j-m}[j_2+m_2]_{m}
\label{eq:ttstar}
\end{eqnarray}}where the generalized hypergeometric function ${}_{4}F_{3}$ is evaluated at $z=1$. Moreover, if $|\lambda|\in\mathbb{N}_+$ the lone-star product corresponds to associative matrix multiplication compatible with $SU(N)$, being $N=|\lambda|$, and the limit $\lambda\to\infty$ leads to $\mathfrak{s}\mathfrak{v}\mathfrak{e}\mathfrak{c}\mathfrak{t}(S^2)$.
Using this product, one obtains the deformed commutator as
\begin{eqnarray}
[T^{j_1}_{m_1},T^{j_2}_{m_2}]=T^{j_1}_{m_1}\star T^{j_{2}}_{m_2}-T^{j_2}_{m_2}\star T^{j_{1}}_{m_1}\,.
\label{eq:com}
\end{eqnarray}
Replacing (\ref{eq:ttcom}) by (\ref{eq:ttstar}) and (\ref{eq:com}), combined with (\ref{eq:tscom}) and (\ref{eq:sscom}), we may investigate whether $hs[\lambda]$ extends to a linear deformation of the entire $\mathfrak{v}\mathfrak{e}\mathfrak{c}\mathfrak{t}(S^2)$ algebra. Imposing the same conditions and following an analogous procedure as before, we find that the Jacobi identity $[T^2_{-2}, T^{3}_0,T^1_{-1}]$  cannot be verified, leading to the conclusion: 
\begin{nohslambdaa} 
\begin{itemize}
\item[] ~\\ The one parameter deformation of the algebra of area-preserving diffeomorphisms on the sphere, $hs[\lambda]$, does not extend to the full algebra of smooth diffeomorphisms on the sphere, $\mathfrak{v}\mathfrak{e}\mathfrak{c}\mathfrak{t}(S^2)$, if the Jacobi identities, parity and vector representation of the generators under rotations are satisfied \footref{reqs}.
\end{itemize}
\label{theorem:nohslambda}
\end{nohslambdaa}
\noindent This result agrees and provides further support for the previous proposal on the absence of linear deformations of $\mathfrak{v}\mathfrak{e}\mathfrak{c}\mathfrak{t}(S^2)$.

\subsubsection{Deformations of the chiral subalgebra}
While $\mathfrak{v}\mathfrak{e}\mathfrak{c}\mathfrak{t}(S^2)$ appears to admit no linear deformations, this might still leave room to the possibility that, in addition to  $\mathfrak{s}\mathfrak{v}\mathfrak{e}\mathfrak{c}\mathfrak{t}(S^2)$, other subalgebras admit linear deformations. With this in mind, we now consider possible linear deformations of the chiral subalgebras $A^{\pm}$. In this case, the assumptions we adopt are:
\begin{enumerate}
    \item The Jacobi identities have to be satisfied.
    \item The generators $(A^j_m)^{\pm}$ have to transform as spherical tensors under $T^1_m$.
    \item The possible deformations have to include the non-deformed classical chiral subalgebras.
\end{enumerate}
The second requirement leads to the ansatz
\begin{align}
[A^{j_1}_{m_1}, A^{j_{2}}_{m_2}]=&\sum^{j_{1}+j_{2}}_{j=|j_1-j_2|}G(j_1,j_2,j)A^{j}_{m_1+m_2}\times C^{j_1j_2j}_{m_1m_2m} \label{eq:aacom}
\end{align}
with 
{\footnotesize
$$C^{j_1j_2j}_{m_1m_2m}=\sum^{j_1+j_2-j}_{m=0} \binom{j_1+j_2-j}{m}[j_1-m_1]_m[j_1+m_1]_{j_1+j_2-j-m}[j_2-m_2]_{j_1+j_2-j-m}[j_2+m_2]_{m}.$$}Performing an analysis identical to the previous cases up to $j_{\text{max}}=7$, we observe that the algebra of commutators with $j\leq 3$ is completely determined. It is again clearly noticeable that the number of independent equations grows faster than the number of free coefficients, and from the computational perspective it seems clear that the algebra will not admit deformations at higher $j$. Thus, we collect strong evidence in favor of:

\begin{nochiraldef} 

\begin{itemize}

    \item[] ~\\ The chiral subalgebras of smooth diffeomorphisms on the sphere, $A^{\pm}$, do not admit linear deformations satisfying the Jacobi identities, and vector representation of the generators under rotations.
\end{itemize}

\label{theorem:nodefa}
\end{nochiraldef}

\subsubsection*{Discussion} 

Let us briefly analyze the main result obtained in this section and its implications. We have gathered strong evidence in favor of a no-go theorem for deformations of $\mathfrak{v}\mathfrak{e}\mathfrak{c}\mathfrak{t}(S^2)$ under the following assumptions: the deformation is linear, the Jacobi identities have to be satisfied and the generators have to transform as spherical tensors under $T^1_m$. The last two conditions are necessary if we aim to obtain deformations which are Lie algebras and whose isometry group is still that of the sphere, allowing us to use the Wigner-Eckart theorem in the ansatz \eqref{eq:ttcom}-\eqref{eq:sscom}. We cannot exclude non-linear deformations akin to $W$-algebras \cite{deBoer:1992sy} in the case of $\mathfrak{v}\mathfrak{e}\mathfrak{c}\mathfrak{t}(S^1)$, the  Witt algebra. To our knowledge, such non-linear deformations have not been explored even for the area preserving subalgebra $\mathfrak{svect}(S^2)$. Linearity could be relaxed at the cost of the ansatz \eqref{eq:ttcom}-\eqref{eq:sscom} having to be modified in order to include terms non-linear in the generators in the RHS. This analysis would be certainly more involved, as we would have to figure out how to efficiently use the Wigner-Eckart theorem and the conservation of angular momentum to constrain the allowed non-linearities. Besides, the computational power required to perform such an analysis is beyond our present capacities. Nevertheless, such an  exploration is definitely worth to be pursued in future studies.

Most of the implications of this result are surely yet to be unveiled, even though we can already notice some important consequences. Firstly, the rigidity under linear deformations of $\mathfrak{v}\mathfrak{e}\mathfrak{c}\mathfrak{t}(S^2)$ is in sharp contrast with the well-known $hs[\lambda]$ deformation of $\mathfrak{svect}(S^2)$. The latter reduces to $SU(N)$ for integer $\lambda$ and defines a large $N$ discretization of $\mathfrak{svect}(S^2)$ \cite{Pope:1990kc}, which has been linked to membrane quantization \cite{1982PhDT........32H,deWit:1988wri,Nicolai:1998ic}. The possibility of a similar discretization for $\mathfrak{v}\mathfrak{e}\mathfrak{c}\mathfrak{t}(S^2)$ is ruled out by our analysis, at least at a linear level, which points towards a fundamental difference between algebras of diffeomorphism and their area preserving subalgebras. Furthermore, we expect the rigidity of $\mathfrak{v}\mathfrak{e}\mathfrak{c}\mathfrak{t}(S^2)$ to play a fundamental role in the understanding of its potential representations and (quantum) deformations. For instance, our result might well pose constraints to generalize the quantum deformations of the $\mathfrak{bms}$ algebra, studied in \cite{Borowiec:2020ddg}, to $\mathfrak{gbms}$ \cite{Campiglia:2014yka,Campiglia:2015yka}, $\mathfrak{gbms}_s$ \cite{Enriquez-Rojo:2021blc} and $\mathfrak{bmsw}$ \cite{Freidel:2021yqe}, which are non-central extensions of $\mathfrak{v}\mathfrak{e}\mathfrak{c}\mathfrak{t}(S^2)$ arising in the study of asymptotically flat and FLRW spacetimes. 

It would certainly be interesting to explore whether the rigidity of $\mathfrak{vect}(S^2)$ extends to other two-dimensional surfaces like the plane or the torus. As far as we are aware, such analysis have not been performed so far in the literature. Unfortunately, our algorithm does not straightforwardly extend to theses spaces. The main obstacle is the lack of spherical symmetry organizing the generators and the subsequent loss of the Wigner-Eckart theorem, which severely constrains the free coefficients to be determined in \eqref{eq:ttcom}-\eqref{eq:sscom}. As a consequence, the number of free coefficients grows substantially making it very challenging to constrain them efficiently.

As a final comment, let us note that, while $\mathfrak{v}\mathfrak{e}\mathfrak{c}\mathfrak{t}(S^1)$ admits a central extension, the Virasoro algebra, it was known for a long time \cite{Bars:1988uj} that $\mathfrak{v}\mathfrak{e}\mathfrak{c}\mathfrak{t}(S^2)$ does not admit central extensions\footnote{More precisely, there the absence of central extensions for area preserving diffeomorphisms was shown. But that is sufficient to imply the result.}. On the other hand, non-central extensions do exist but their description is cumbersome due to the complicated form of the structure constants of  $\mathfrak{v}\mathfrak{e}\mathfrak{c}\mathfrak{t}(S^2)$. 
In the next section we will discuss some extensions by embedding $\mathfrak{v}\mathfrak{e}\mathfrak{c}\mathfrak{t}(S^2)$ in $\mathfrak{v}\mathfrak{e}\mathfrak{c}\mathfrak{t}(\CC^*)$.

\section{Embedding in  \(\mathfrak{v}\mathfrak{e}\mathfrak{c}\mathfrak{t}(\CC^*)\)}
\label{lvectS2}
We can embed $\mathfrak{v}\mathfrak{e}\mathfrak{c}\mathfrak{t}(S^2)$ in $\mathfrak{v}\mathfrak{e}\mathfrak{c}\mathfrak{t}(\mathbb{C})$ simply by replacing (\ref{eq:agen}) by arbitrary smooth holomorphic and anti-holomorphic vector fields on $\CC$. More generally, if we allow the vector fields to be singular at the origin, we can choose the following basis of $\mathfrak{v}\mathfrak{e}\mathfrak{c}\mathfrak{t}(\CC^*)$   
\begin{equation}
 \mathcal{L}_{m,n}=-z^{m+1}\bar{z}^n\partial_z \ , \ \ \hat{\mathcal{L}}_{m,n}=-z^{m}\bar{z}^{n+1}\partial_{\bar{z}} \ , \label{eq:localgen}
\end{equation}
with $m,n\in\mathbb{Z}$ and non-vanishing commutators 
\begin{align}
    [\mathcal{L}_{m,n},\mathcal{L}_{r,s}]=(m-r)\mathcal{L}_{m+r,n+s} \ ,  \ \  [\hat{\mathcal{L}}_{m,n},\hat{\mathcal{L}}_{r,s}]=(n-s)\hat{\mathcal{L}}_{m+r,n+s} \ , \label{eq:llhat} \\ 
    [\mathcal{L}_{m,n},\hat{\mathcal{L}}_{r,s}]=-r\hat{\mathcal{L}}_{m+r,n+s}+n\mathcal{L}_{m+r,n+s} \ \label{eq:mixed}.\nonumber
\end{align}
In fact, (\ref{eq:llhat}) makes it clear that (\ref{eq:localgen}) is isomorphic to  $\mathfrak{v}\mathfrak{e}\mathfrak{c}\mathfrak{t}(\CC^*)$ and to $\mathfrak{v}\mathfrak{e}\mathfrak{c}\mathfrak{t}(\mathbb{T}^2)$ (see \cite{Larsson:1991xh,Larsson:1992fx,Larsson:1992bt,moody,Larsson:1997dt,alberte:2010,Larsson:2015sla} for a detailed analysis and some representations). This is not surprising since both can be obtained from the two-punctured sphere with suitable identifications. Thus we actually have 
\begin{align}
    \mathfrak{v}\mathfrak{e}\mathfrak{c}\mathfrak{t}(S^2)\hookrightarrow \mathfrak{v}\mathfrak{e}\mathfrak{c}\mathfrak{t}(\CC^*)\hookleftarrow \mathfrak{v}\mathfrak{e}\mathfrak{c}\mathfrak{t}(\mathbb{T}^2)
\end{align}
which is compatible with the geometric picture of the cylinder being an open subset either to $S^2$ or $\mathbb{T}^2$.

Unlike $(A^l_m)^{\pm}$, the basis (\ref{eq:localgen}) does not diagonalize the $so(3)$ Casimir\footnote{As it is well known from textbook literature on the Runge-Lenz vector.} but, instead, simultaneously diagonalizes 
\begin{align}
    (A^1_0)^\pm\;\text{and}\; (A^\pm)^2 \ .
\end{align}
As already mentioned, these vector fields are generally singular on $S^2$, for $z,\bar{z}\to0$ and $z,\bar{z}\to\infty$. In fact, they form an over-complete basis for the global vector fields in $\mathfrak{v}\mathfrak{e}\mathfrak{c}\mathfrak{t}(S^2)$.  
This can be seen by noticing that the global vector fields on $S^2$ have the form
\begin{align}
    \frac{1}{(1+z\bar{z})^{l-1}}P(z^{a}\bar{z}^{b})\partial_{z} \quad \text{and}\quad
    \frac{1}{(1+z\bar{z})^{l-1}} P(z^{a}\bar{z}^{b})\partial_{\bar{z}} \ , 
    \label{PP}
\end{align}
where $P(z^{a}\bar{z}^{b})$ is a polynomial. Thus expanding around the south pole  ($z\bar{z}\to0$)
\begin{equation}
    \frac{1}{1+z\bar{z}}=1-z\bar{z}+(z\bar{z})^2-...=\sum_{p\geq0}(-1)^{p}(z\bar{z})^{p} \ , \nonumber
\end{equation}
or, around the north pole  ($z\bar{z}\to\infty$)
\begin{equation}
    \frac{1}{1+z\bar{z}}=(z\bar{z})^{-1}(1-(z\bar{z})^{-1}+(z\bar{z})^{-2})=\sum_{p\geq0}(-1)^{p}(z\bar{z})^{-(p+1)} \ , \nonumber
\end{equation}
they are clearly infinite linear combinations of (\ref{eq:localgen}). 

We see that this gives two different ways of representing a smooth vector field on $S^2$ as an infinite linear combination of the elements (\ref{eq:localgen}). This is analogous to the fact that the Taylor expansion of a rational function of two variables $z$ and $w$ in two different regions ($|z|\gg|w|$ and $|w|\gg|z|$) leads to different formal power
series representing the function \cite{Frenkel:2004jn}.

\subsection{Extensions of $\mathfrak{v}\mathfrak{e}\mathfrak{c}\mathfrak{t}(\CC^*)$ }
As shown in \cite{Larsson:1992fx,moody}, $\mathfrak{v}\mathfrak{e}\mathfrak{c}\mathfrak{t}(\CC^*)$ and $\mathfrak{v}\mathfrak{e}\mathfrak{c}\mathfrak{t}(\mathbb{T}^2)$ do not admit central extensions either. However, there are non-central extensions which reduce to the Virasoro central extension when viewed as a subalgebra \cite{Larsson:1992fx,moody}. They can be described as 
\begin{align}
    [\mathcal{L}_{m,n},\mathcal{L}_{r,s}]&=(m-r)\mathcal{L}_{m+r,n+s}-mr(c_1+c_2)(m\mathcal{S}_{m+r,n+s}+n\hat{\mathcal{S}}_{m+r,n+s}) \ , \nonumber\\ [\hat{\mathcal{L}}_{m,n},\hat{\mathcal{L}}_{r,s}]&=(n-s)\hat{\mathcal{L}}_{m+r,n+s}-ns(c_1+c_2)(m\mathcal{S}_{m+r,n+s}+n\hat{\mathcal{S}}_{m+r,n+s}) \ , \nonumber \\ 
    [\mathcal{L}_{m,n},\hat{\mathcal{L}}_{r,s}]&=-r\hat{\mathcal{L}}_{m+r,n+s}+n\mathcal{L}_{m+r,n+s}-(c_1nr+c_2ms)(m\mathcal{S}_{m+r,n+s}+n\hat{\mathcal{S}}_{m+r,n+s}) \ , \nonumber \\
    [\mathcal{L}_{m,n},\mathcal{S}_{r,s}]&=s\hat{\mathcal{S}}_{m+r,n+s} \ , \ \ \ 
    [\mathcal{L}_{m,n},\hat{\mathcal{S}}_{r,s}]=-r\hat{\mathcal{S}}_{m+r,n+s} \ , \nonumber \\
    [\hat{\mathcal{L}}_{m,n},\mathcal{S}_{r,s}]&=-s\mathcal{S}_{m+r,n+s} \ , \ \ \ 
    [\hat{\mathcal{L}}_{m,n},\hat{\mathcal{S}}_{r,s}]=r\mathcal{S}_{m+r,n+s} \ ,\label{llc}
\end{align}
subject to $m\mathcal{S}_{m,n}+n\hat{\mathcal{S}}_{m,n}=0$. It is not hard to see that the non-central extensions parametrized by $c_1$ and $c_2$ are not compatible with regularity at the origin: Comparing (\ref{PP}) with (\ref{eq:localgen}), we see that $\mathfrak{v}\mathfrak{e}\mathfrak{c}\mathfrak{t}(\mathbb{C})$ contains the elements 
\begin{equation}
 \mathcal{L}_{m\geq -1,n\geq 0}
\quad \text{and}\quad \hat{\mathcal{L}}_{m\geq 0,n\geq -1}\ . \label{ES2}
\end{equation}
Then, for non-vanishing $c_1$ and $c_2$ (\ref{llc}) contains non-central extensions $\mathcal{S}_{r,s}$ and $\hat{\mathcal{S}}_{r,s}$ with arbitrary negative values of $r$ and $s$. 

\subsubsection{$gW(a,b,\bar{a})$ - a family of non-central extensions} 

Another class of non-central extensions is obtained by considering representations of $\mathfrak{v}\mathfrak{e}\mathfrak{c}\mathfrak{t}(\CC^*)$ on tensors. For instance, the extension 
\begin{align}
    [\mathcal{L}_{m,n},\mathcal{L}_{r,s}]=(m-r)\mathcal{L}_{m+r,n+s} \ , \ \ [\hat{\mathcal{L}}_{m,n},\hat{\mathcal{L}}_{r,s}]=(n-s)\hat{\mathcal{L}}_{m+r,n+s} \ , \nonumber \\ 
    [\mathcal{L}_{m,n},\hat{\mathcal{L}}_{r,s}]=-r\hat{\mathcal{L}}_{m+r,n+s}+n\mathcal{L}_{m+r,n+s} \ , \nonumber \\
    [\mathcal{L}_{m,n},T_{pq}]=\left[\frac{(m+1)}{2}(1+s)-p \right]T_{p+m,q+n}  \ , \nonumber \\
    [\hat{\mathcal{L}}_{m,n},T_{pq}]=\left[\frac{(n+1)}{2}(1+s)-q \right]T_{p+m,q+n} \ ,
\end{align}
known as $\mathfrak{g}\mathfrak{b}\mathfrak{m}\mathfrak{s}_s\simeq\mathfrak{v}\mathfrak{e}\mathfrak{c}\mathfrak{t}(\CC^*)\ltimes_s\mathfrak{s}_s$ algebras\footnote{That is generalised $\mathfrak{b}\mathfrak{m}\mathfrak{s}_s$,  where $\mathfrak{s}_s$ stands for conformally weighted supertranslations.} \cite{Enriquez-Rojo:2021blc} has recently played a role as an asymptotic symmetry algebra of decelerating asymptotically spatially flat Friedmann spacetimes at $\mathcal{I}^+$ \cite{Bonga:2020fhx,Rojo:2020zlz,Enriquez-Rojo:2021blc}. It turns out that this algebra forms part of a bigger family of deformations of $\mathfrak{g}\mathfrak{b}\mathfrak{m}\mathfrak{s}$, analogously to the $W(a,b;\bar{a},\bar{b})$ deformations for $\mathfrak{b}\mathfrak{m}\mathfrak{s}$ \cite{Safari:2019zmc,Safari:2020pje} \footnote{In fact, the $W(a,b;\bar{a},\bar{b})$ algebras are given by (\ref{eq:gW1})-(\ref{eq:gW4}) if we restrict to $\mathcal{L}_{m,0}$ and $\hat{\mathcal{L}}_{0,n}$.}. Let us denote these algebras by generalized $W(a,b;\bar{a},\bar{b})$, or $gW(a,b;\bar{a},\bar{b})$, and postulate the commutation relations as
\begin{align}
    [\mathcal{L}_{m,n},\mathcal{L}_{r,s}]=(m-r)\mathcal{L}_{m+r,n+s} \ , \ \ [\hat{\mathcal{L}}_{m,n},\hat{\mathcal{L}}_{r,s}]=(n-s)\hat{\mathcal{L}}_{m+r,n+s} \ , \label{eq:gW1}\\ 
    [\mathcal{L}_{m,n},\hat{\mathcal{L}}_{r,s}]=-r\hat{\mathcal{L}}_{m+r,n+s}+n\mathcal{L}_{m+r,n+s} \ , \label{eq:gW2}\\
    [\mathcal{L}_{m,n},T_{pq}]=-\left[p+bm+a \right]T_{p+m,q+n}  \ , \label{eq:gW3}\\
    [\hat{\mathcal{L}}_{m,n},T_{pq}]=-\left[q+\bar{b}n+\bar{a} \right]T_{p+m,q+n} \ . \label{eq:gW4}
\end{align}
With this general ansatz these algebras are actually  inconsistent due to the Jacobi identity
\begin{align}
    [\mathcal{L}_{m,n},[\hat{\mathcal{L}}_{r,s},T_{pq}]]+ \text{cyclic permutations} = nr(b-\bar{b})\overset{!}{=}0 \ .
\end{align}
Consequently, we find that, unless $n=0$ and/or $r=0$, which correspond to the Witt subalgebras, we are forced to  to set $b=\bar{b}$. Therefore, the family of algebras is actually $gW(a,b;\bar{a})$. Some examples of algebras in this family are given by $\mathfrak{g}\mathfrak{b}\mathfrak{m}\mathfrak{s}\simeq gW(-\frac{1}{2},-\frac{1}{2};-\frac{1}{2})$ and $\mathfrak{g}\mathfrak{b}\mathfrak{m}\mathfrak{s}_s\simeq gW(-\frac{1+s}{2},-\frac{1+s}{2};-\frac{1+s}{2})$. We note in passing that if the superrotation-like vector fields appearing in the near horizon symmetry algebras described in \cite{Donnay:2015abr,Donnay:2019zif} and \cite{Grumiller:2019fmp} are not constrained to satisfy the conformal Killing equation, the latter are described by $gW(0,0;0)$ and $gW(a,a;a)$  respectively.

\subsubsection*{Free field realization}

It might seem difficult to find a representation of these complicated algebras. Nevertheless, it turns out that there exists a Heisenberg-like construction which provides us with a free field realization for the family $gW(a,b;\bar{a})$. This is given by
\begin{align}
    \mathcal{L}_{m,n}&=\sum_{\alpha,\beta}(\alpha+(b-1)m+a)\bar{a}_{m-\alpha,n-\beta}a_{\alpha,\beta} \label{eq:firstlocS2} \ , \\
    \hat{\mathcal{L}}_{m,n}&=\sum_{\alpha,\beta}(\beta+(\bar{b}-1)n+\bar{a})\bar{a}_{m-\alpha,n-\beta}a_{\alpha,\beta} \label{eq:seclocS2} \ , \\
    T_{p,q}&=a_{p,q} \ , \label{eq:thirdlocS2}
\end{align}
with 
\begin{align}
    [a_{\alpha,\beta},\bar{a}_{\gamma,\delta}]= \delta_{\alpha+\gamma,0}\delta_{\beta+\delta,0} 
\end{align}
and $b=\bar{b}$.

This free field realization helps us to visualize the physical meaning of the uniparametric family of deformations $\mathfrak{g}\mathfrak{b}\mathfrak{m}\mathfrak{s}_s\simeq gW(-\frac{1+s}{2},-\frac{1+s}{2};-\frac{1+s}{2})$, being $s$ related to the weight in the lattice.  Besides, it is evident that this representation describes also the subfamily $W(a,b;\bar{a},\bar{b})$ with $n=0$ in (\ref{eq:firstlocS2}) and $m=0$ in (\ref{eq:seclocS2}) and sheds light on the symmetries of the coefficients $a,\bar{a},b,\bar{b}$ described in section 5.3 of \cite{Safari:2020pje}. 

\subsubsection{$\hat{gW}(a,b,\bar{a})$ - two compatible non-central extensions}

Guided by the fact that $W(a,b;\bar{a},\bar{b})$ admits a central extension, $\hat{W}(a,b;\bar{a},\bar{b})$, obtained by adding central extensions to both Witt subalgebras, we expect to find an equivalent extension for $gW(a,b,\bar{a})$, which we will call $\hat{gW}(a,b,\bar{a})$, as the addition of both non-central extensions of $\mathfrak{v}\mathfrak{e}\mathfrak{c}\mathfrak{t}(\CC^*)$
\begin{align}
    [\mathcal{L}_{m,n},\mathcal{L}_{r,s}]&=(m-r)\mathcal{L}_{m+r,n+s}-mr(c_1+c_2)(m\mathcal{S}_{m+r,n+s}+n\hat{\mathcal{S}}_{m+r,n+s}) \ , \nonumber\\ [\hat{\mathcal{L}}_{m,n},\hat{\mathcal{L}}_{r,s}]&=(n-s)\hat{\mathcal{L}}_{m+r,n+s}-ns(c_1+c_2)(m\mathcal{S}_{m+r,n+s}+n\hat{\mathcal{S}}_{m+r,n+s}) \ , \nonumber \\ 
    [\mathcal{L}_{m,n},\hat{\mathcal{L}}_{r,s}]&=-r\hat{\mathcal{L}}_{m+r,n+s}+n\mathcal{L}_{m+r,n+s}-(c_1nr+c_2ms)(m\mathcal{S}_{m+r,n+s}+n\hat{\mathcal{S}}_{m+r,n+s}) \ , \nonumber \\
    [\mathcal{L}_{m,n},\mathcal{S}_{r,s}]&=s\hat{\mathcal{S}}_{m+r,n+s} \ , \ \ \ 
    [\mathcal{L}_{m,n},\hat{\mathcal{S}}_{r,s}]=-r\hat{\mathcal{S}}_{m+r,n+s} \ , \nonumber \\
    [\hat{\mathcal{L}}_{m,n},\mathcal{S}_{r,s}]&=-s\mathcal{S}_{m+r,n+s} \ , \ \ \ 
    [\hat{\mathcal{L}}_{m,n},\hat{\mathcal{S}}_{r,s}]=r\mathcal{S}_{m+r,n+s} \ , \label{nct} \\
    [\mathcal{L}_{m,n},T_{pq}]&=-\left[p+bm+a \right]T_{p+m,q+n}  \ ,  \ \
    [\hat{\mathcal{L}}_{m,n},T_{pq}]=-\left[q+bn+\bar{a} \right]T_{p+m,q+n} \ . \nonumber
\end{align}
It turns out that both non-central extensions are compatible, in terms of Jacobi identities, if the commutators among their generators vanish, $[T_{pq},\mathcal{S}_{r,s}]=[T_{pq},\hat{\mathcal{S}}_{r,s}]=0$. Of course, we cannot interpret this algebra as an extension of $\mathfrak{v}\mathfrak{e}\mathfrak{c}\mathfrak{t}(S^2)$ since, as mentioned above, $c_1=c_2=0$ for the latter \cite{Bars:1988uj}. The same happens to the centrally extended $\hat{W}(a,b;\bar{a},\bar{b})$ algebras, although one can define them abstractly using the punctured complex plane as suggested by the works of \cite{Barnich:2017ubf,Ball:2019atb,Safari:2019zmc,Safari:2020pje,Barnich:2021dta}. In fact, $\hat{W}(a,b;\bar{a},\bar{b})$ is a subfamily of (\ref{nct}) after using the conditions
\begin{align}
  m\mathcal{S}_{m,n}+n\hat{\mathcal{S}}_{m,n}=0 \ , \ \ \ \mathcal{S}_{m,0}=\mathcal{S}_{0,0}\delta_{m0} \ , \ \ \ \hat{\mathcal{S}}_{0,n}=\hat{\mathcal{S}}_{0,0}\delta_{n0} \ , \nonumber \\
   (c_1+c_2)\mathcal{S}_{0,0}=\frac{c}{12} \ , \ \ \ \ (c_1+c_2)\hat{\mathcal{S}}_{0,0}=\frac{\bar{c}}{12} \ , 
\end{align}
which allow to recover Virasoro central extensions in the one-dimensional limit.

\section{Summary and conclusions}
\label{conclusion}

The object of study of this paper is the algebra of vector fields on the sphere. Besides being mathematically interesting per se and scarcely studied, it pops up ubiquitously in physics literature. Although $\text{Diff}(S^2)$ plays a major role in membrane theory \cite{1982PhDT........32H,deWit:1988wri,Nicolai:1998ic} and fluid-gravity duality \cite{Penna:2015gza,Penna:2017bdn}, our main motivation emerges from recent investigations in asymptotically flat \cite{Campiglia:2014yka,Campiglia:2015yka,Donnelly:2020xgu,Freidel:2021yqe,Freidel:2021cbc} and asymptotically spatially flat FLRW \cite{Bonga:2020fhx,Rojo:2020zlz,Enriquez-Rojo:2021blc} spacetimes, where the asymptotic symmetry algebras contain as superrotation subalgebra that of vector fields on $S^2$. 

In section \ref{vectS2}, we restricted to smooth vector fields, which form the algebra $\mathfrak{v}\mathfrak{e}\mathfrak{c}\mathfrak{t}(S^2)$. Firstly, we described the structure of this algebra in the conventional area preserving ($T^l_m$) and non-area preserving ($S^l_m$) vector fields, which we used to investigate possible linear deformations of $\mathfrak{v}\mathfrak{e}\mathfrak{c}\mathfrak{t}(S^2)$ with the help of an attached Mathematica file \textsc{vectS2deformations.nb} \cite{Mathematica}, where explicit details on the computations can be found. Next, with the help of stereographic coordinates, we found a more illuminating chiral basis that splits into vector fields with purely holomorphic $(A^l_m)^+$ and antiholomorphic components $(A^l_m)^-$. 

In section \ref{lvectS2}, we loosened the smoothness condition for the vector fields and embedded $\mathfrak{v}\mathfrak{e}\mathfrak{c}\mathfrak{t}(S^2)$ in $\mathfrak{v}\mathfrak{e}\mathfrak{c}\mathfrak{t}(\CC^*)$, allowing for two punctures. In terms of it, we examined physically relevant non-central extensions and came up with some simple free field realizations. Remarkably, the two-punctured Riemann sphere, where the conformal subalgebra of (\ref{eq:llhat}) and of (\ref{llc}) is consistently realized, has been argued to be the relevant one for celestial scattering amplitudes and soft theorems in the context of $\mathfrak{b}\mathfrak{m}\mathfrak{s}$ \cite{Strominger:2016wns,Barnich:2017ubf,Ball:2019atb,Barnich:2021dta}. Analogously, we expect the complete (\ref{eq:llhat}) and (\ref{llc}) to play the equivalent role for $\mathfrak{g}\mathfrak{b}\mathfrak{m}\mathfrak{s}$.

Let us recall our most important results: 

\begin{itemize}
    \item By means of the chiral basis, we observed that $(A^l_{\pm})^{+}$ and $(A^l_{\pm})^{-}$ describe half-Witt subalgebras generated by smooth vector fields on $S^2$. Moreover, both chiral subalgebras $A^{\pm}$ can be reconstructed from a half-Witt subalgebra and the action of rotation operators as described in the picture \ref{Agen}. We find the chiral basis especially illuminating and hope that it will help in future studies of this important algebra.
    
    \item We found that the Jacobi identities fix the structure constants for small values of $j$ completely, which suggests that $\mathfrak{v}\mathfrak{e}\mathfrak{c}\mathfrak{t}(S^2)$ does not admit linear deformations satisfying Jacobi identities, being compatible with parity and transforming in given representations of the rotation group. In particular, we showed that the higher-spin one parameter deformation of $\mathfrak{s}\mathfrak{v}\mathfrak{e}\mathfrak{c}\mathfrak{t}(S^2)$, $hs[\lambda]$ \cite{Pope:1990kc}, does not extend to $\mathfrak{v}\mathfrak{e}\mathfrak{c}\mathfrak{t}(S^2)$ under these requirements. For the chiral subalgebras of $\mathfrak{v}\mathfrak{e}\mathfrak{c}\mathfrak{t}(S^2)$, $A^{\pm}$ we find by the same method that it should not  admit linear deformations satisfying Jacobi identities and vector representation of the generators under rotations. 
   
    \item In terms of the locally defined vector fields on the two-punctured sphere, we uncovered a three parameter family of non-central extensions $gW(a,b,\bar{a})$ which contains the asymptotic symmetry algebra of asymptotically flat ($\mathfrak{g}\mathfrak{b}\mathfrak{m}\mathfrak{s}$ \cite{Campiglia:2014yka,Campiglia:2015yka}) and asymptotically decelerating spatially flat FLRW ($\mathfrak{g}\mathfrak{b}\mathfrak{m}\mathfrak{s}_s$ \cite{Enriquez-Rojo:2021blc}) spacetimes at future null infinity. It generalizes and contains the $W(a,b;\bar{a},\bar{b})$ family of deformations for $\mathfrak{b}\mathfrak{m}\mathfrak{s}$ \cite{Safari:2019zmc,Safari:2020pje} and admits a simple free field realization compatible with the ones described in \cite{Larsson:1991xh,Larsson:1992bt}. In addition, guided by the fact that $W(a,b;\bar{a},\bar{b})$ admits a central extension, $\hat{W}(a,b;\bar{a},\bar{b})$, obtained by centrally extending both Witt algebras, we found an equivalent extension for $gW(a,b,\bar{a})$, which we denoted by $\hat{gW}(a,b,\bar{a})$. 
\end{itemize}

Finally, we briefly list some open questions and especially interesting research directions. 

\begin{itemize}
    \item It was shown in \cite{Bars:1988uj} that $\mathfrak{s}\mathfrak{v}\mathfrak{e}\mathfrak{c}\mathfrak{t}(S^2)$ does not admit any central extension and our results strongly suggest that the complete algebra $\mathfrak{v}\mathfrak{e}\mathfrak{c}\mathfrak{t}(S^2)$ does not admit linear deformations. It would be certainly interesting to explore whether or not $\mathfrak{v}\mathfrak{e}\mathfrak{c}\mathfrak{t}(S^2)$ admits non-central extensions and/or non-linear deformations \cite{deBoer:1992sy}.  
    
    \item Locally, Witt and Virasoro algebras have been shown to not admit linear deformations \cite{fia,schlichenmaier}, although they allow for non-linear ones \cite{Curtright:1989sw,doi:10.1063/1.530518}. Taking into account that $\mathfrak{v}\mathfrak{e}\mathfrak{c}\mathfrak{t}(\CC^*)\hookleftarrow \mathfrak{v}\mathfrak{e}\mathfrak{c}\mathfrak{t}(\mathbb{T}^2)$ is its more direct two dimensional generalization, it would be desirable to investigate its possible linear and non-linear deformations. In particular, $\mathfrak{s}\mathfrak{v}\mathfrak{e}\mathfrak{c}\mathfrak{t}(\mathbb{T}^2)$ admits a one parameter deformation similar to $hs[\lambda]$ \cite{Pope:1989cr} which might or not extend to $\mathfrak{v}\mathfrak{e}\mathfrak{c}\mathfrak{t}(\mathbb{T}^2)$.
    
    \item We did not dig into field realizations of $\mathfrak{v}\mathfrak{e}\mathfrak{c}\mathfrak{t}(S^2)$ and $\hat{gW}(a,b,\bar{a})$ but it would be appealing to explore them in detail.  
    
    \item Recently, a double non-central extension of $\mathfrak{v}\mathfrak{e}\mathfrak{c}\mathfrak{t}(S^2)$, called Weyl $\mathfrak{b}\mathfrak{m}\mathfrak{s}$ ($\mathfrak{b}\mathfrak{m}\mathfrak{s}\mathfrak{w}$), has been proposed in \cite{Freidel:2021yqe} to be the most general extension of the $\mathfrak{b}\mathfrak{m}\mathfrak{s}$ algebra \footnote{Closely related symmetry algebras analyzing the inclusion of Weyl scaling have been previously discussed in \cite{Adami:2020amw,Adami:2020ugu,Adami:2021sko}.}. It would be very interesting to generalize this construction to asymptotically FLRW spacetimes, to study its family of deformations and to analyze possible field realizations in a similar way we described in this paper for the other non-central extensions. Similar considerations apply for the so-called corner symmetry and extended corner symmetry algebras \cite{Donnelly:2020xgu,Freidel:2021cbc}, which also non-centrally extend $\mathfrak{v}\mathfrak{e}\mathfrak{c}\mathfrak{t}(S^2)$.
    
\end{itemize}

\vspace{1em}
\section*{Acknowledgements}
We would like to thank T.~Heckelbacher for very helpful feedback on the manuscript and I.~Kharag for proofreading this paper. This work was funded by the Excellence Cluster Origins of the DFG under Germany’s Excellence Strategy EXC-2094 390783311 and the Grant Agency of the Czech Republic under the grant EXPRO 20-25775X.



\clearpage
\nocite{*}
\bibliography{references2}
\bibliographystyle{JHEP}


\end{document}